\documentclass{cjpsuppl}
 \usepackage{amsbsy,amssymb,graphicx} 
 \usepackage{cite}

\def\ppbar{{\mbox{p}\bar{\mbox{p}}}}%
\def\pbar{{\bar{\mbox{p}}}}%
\def\lam{{\Lambda}}%
\def\alam{{\bar{\Lambda}}}%
\def\kperp{{\boldsymbol{\kappa}_{\perp}}}%
\def\gevc{{GeV/c}}
\def\gevcc{{GeV/c$^2$}}
\def\mumu{{\mu^+\mu^-}}
\def\rw{\rightarrow }

 \slacs{.6mm}
 \begin{document}
  \title{Spin physics with antiprotons}

 \authori{ASSIA COLLABORATION\\[2mm]M.~Maggiora$^2$,
V.~Abazov$^1$,G.~Alexeev$^1$, M.~Alexeev$^2$, A.~Amoroso$^2$,
N.~Angelov$^1$, S.~Baginyan$^1$, F.~Balestra$^{2}$, V.A.~Baranov$^1$,
Yu.~Batusov$^1$, I.~Belolaptikov$^1$, R.~Bertini$^{2}$,
A.~Bianconi$^3$, R.~Birsa$^{10}$, T.~Blokhintseva$^1$,
A.~Bonyushkina$^1$, F.~Bradamante$^{10}$, A.~Bressan$^{10}$,
M.P.~Bussa$^2$, V.~Butenko$^1$, M.~Colantoni$^4$, M.~Corradini$^{3}$,
S.~Dalla~Torre$^{10}$, A.~Demyanov$^1$, O.~Denisov$^2$,
V.~Drozdov$^1$, J.~Dupak$^8$, G.~Erusalimtsev$^1$, L.~Fava$^4$,
A.~Ferrero$^2$, L.~Ferrero$^2$, M.~Finger,\,Jr.$^{1,5}$, M.~Finger $^6$,
V.~Frolov$^2$, R.~Garfagnini$^2$, M.~Giorgi$^{10}$, O.~Gorchakov$^1$,
A.~Grasso$^2$, V.~Grebenyuk$^1$, V.~Ivanov$^1$, A.~Kalinin$^1$,
V.A~Kalinnikov$^1$, Yu.~Kharzheev$^1$, Yu. Kisselev$^1$,
N.V.~Khomutov$^1$, A.~Kirilov$^1$, E.~Komissarov$^1$,
A.~Kotzinian$^2$, A.S.~Korenchenko$^1$, V.Kovalenko$^1$,
N.P.~Kravchuk$^1$, N.A.~Kuchinski$^1$, E.~Lodi~Rizzini$^{3}$,
V.~Lyashenko$^1$, V.~Malyshev$^1$, A.~Maggiora$^2$, A.~Martin$^{10}$,
Yu.~Merekov$^1$, A.S.~Moiseenko$^1$, A.~Olchevski$^1$,
V.~Panyushkin$^1$, D.~Panzieri$^4$, G.~Piragino$^2$,
G.B.~Pontecorvo$^1$, A.~Popov$^1$, S.~Porokhovoy$^1$,
V.~Pryanichnikov$^1$, M.~Radici$^{11}$, M.P.~Rekalo$^9$, 
A.~Rozhdestvensky$^1$, N.~Russakovich$^1$, P.~Schiavon$^{10}$,
O.~Shevchenko$^1$, A.~Shishkin$^1$, V.A.~Sidorkin$^1$,
N.~Skachkov$^1$, M.~Slunecka$^6$, S.~Sosio$^2$, A.~Srnka$^8$,
V.~Tchalyshev$^1$, F.~Tessarotto$^{10}$, E.~Tomasi$^7$,
F.~Tosello$^2$, E.P.~Velicheva$^1$, L.~Venturelli$^3$,
L.~Vertogradov$^1$, M.~Virius$^8$, G.~Zosi$^2$ and N.~Zurlo$^{3}$}
\addressi{$^1$Dzhelepov Laboratory of Nuclear Problems, JINR, Dubna, Russia\\
$^2$Dipartimento di Fisica ``A. Avogadro'' and INFN - Torino, Italy \\ 
$^3$Universit\`{a} and INFN, Brescia, Italy \\
$^4$Universita' del Piemonte Orientale and INFN sez. di Torino, Italy\\ 
$^5$Czech Technical University , Prague, Czech Republic\\
$^6$Charles University, Prague, Czech Republic\\ 
$^7$DAPNIA,CEN Saclay, France\\ 
$^8$Inst. of Scientific Instruments Academy of Sciences, Brno, Czech Republic \\
$^9$NSC Kharkov Physical Technical Institute, Kharkov, Ukraine \\ 
$^{10}$University of Trieste and INFN Trieste, Italy\\ 
$^{11}$INFN sez. Pavia, Italy}
 \authorii{}    \addressii{}
 \authoriii{}   \addressiii{}
 \authoriv{}    \addressiv{}
 \authorv{}     \addressv{}
 \authorvi{}    \addressvi{}
 \headtitle{Spin physics with antiprotons}
 \headauthor{M.~Maggiora et al.}
 \lastevenhead{M.~Maggiora et al.: Spin physics with antiprotons}
 \pacs{}
 \keywords{spin physics, antiproton, parton distribution functions, transversity}
 \refnum{}
 \daterec{} 
 \suppl{?}  \year{2005} \setcounter{page}{1}
 \maketitle

 \begin{abstract}
 New possibilities arising from the availability at GSI of antiproton
 beams, possibly polarised, are discussed. The investigation of the
 nucleon structure can be boosted by accessing in Drell--Yan processes
 experimental asymmetries related to cross-sections in which the parton
 distribution functions (PDF) only appear, without any contribution from
 fragmentation functions; such processes are not affected by the chiral
 suppression of the transversity function $h_1(x)$. Spin asymmetries in
 hyperon production and Single Spin Asymmetries are discussed as well,
 together with further items like electric and magnetic nucleonic form
 factors and open charm production. Counting rates estimations are
 provided for each physical case. The sketch of a possible
 experimental apparatus is proposed.
 \end{abstract}

 \section{Introduction} 

 The possibility to build a new facility at GSI, a new ring (SIS 300) of
 the Superconducting Synchrotron with a rigidity of 300\,Tm, and to
 extract from it antiprotons, possibly polarised, provides an
 excellent tool to investigate the nucleonic structure. 

 An excellent case to be studied is the di-lepton Drell--Yan production
 where protons and antiprotons annihilate in the initial state. An
 important role may be played also by the evaluation of the spin
 observables in hadron production. I'll concentrate herewith in the
 relevant topics present in the ASSIA LOI \cite{assialoi}.

 All these items can be investigated in the framework of the new GSI
 facility; the key issue is the availability of an antiproton beam,
 with an energy suitable to investigate the parton distribution
 functions in a wide range of the Bjorken kinematic variable x.
 Two different scenarios have been proposed: a
 slow extraction of the antiprotons from the SIS 300 to a both
 longitudinally and transversely polarised proton target; an evolution
 of the GSI HESR toward a collider configuration in which a polarised
 proton beam would collide with an possibly polarised antiproton beam.

\begin{figure}[h] 
 \vspace{-3mm}\bc
 \begin{minipage}[c]{.6\textwidth}
   {\includegraphics[width=\textwidth]{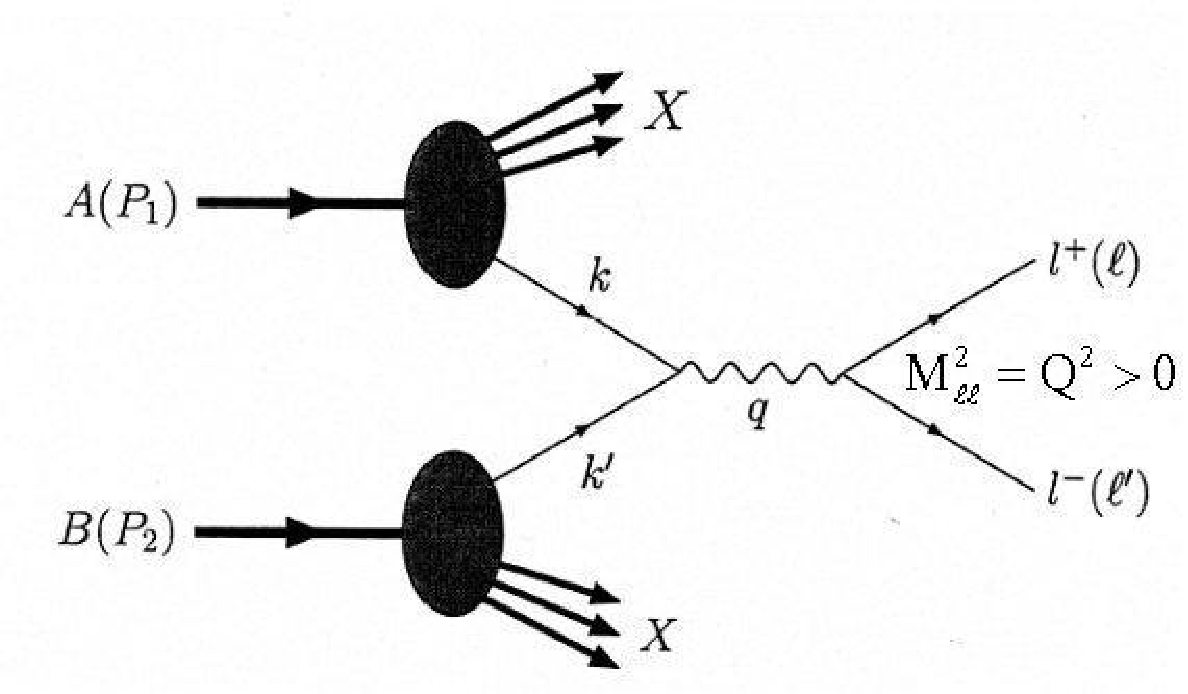}}
 \end{minipage}\hfill
 \begin{minipage}[c]{.39\textwidth}
   {\includegraphics[width=\textwidth]{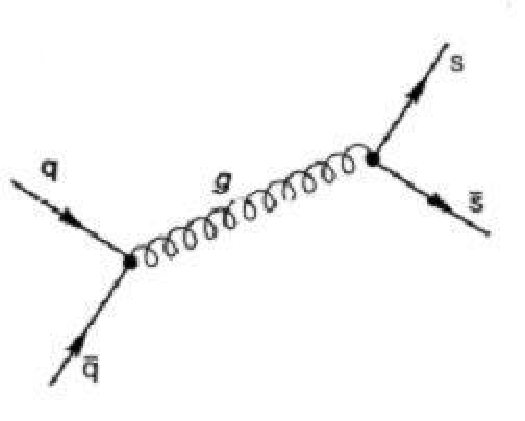}}
 \end{minipage}
 \ec \vspace{-6mm}
 \caption{Left: Drell--Yan dilepton production;
 right: a quark diagram relevant in hyperons production.}
 \label{fig:dy-fey}
 \label{fig:dy-ppss}
 \end{figure}

 The Drell--Yan interactions are known to be affected by low
 cross-sections, and yet the investigation of such processes presents
 several advantages:
 \begin{enumerate}
 \item Due to the non-perturbative vertexes present
 in the Drell--Yan diagrams (Fig.~\ref{fig:dy-fey}~left),
 there is no suppression for chirally odd parton
 distribution functions like transversity.
 \item Other kind of hard processes, like semi-inclusive
 deep-inelastic scattering (SIDIS), can access chirally
 odd distribution functions, but in the case of Drell--Yan
 diagrams the parton distribution functions can be
 directly accessed, while the other processes provide only
 their convolution with unknown polarised quark
 fragmentation functions.
 \item If we use an antiproton probe, all its constituents
 can participate to the reaction; if compared with
 proton-proton or pion-proton scattering, in the
 Drell--Yan process all the partons taking part to the
 reaction can be valence quarks, without the need of sea quark.
 \end{enumerate}

 The selection between  the two scenarios must take into
 account the most relevant parameter, the center of mass
 energy, that must be high enough to span the desired
 kinematic region. A complete experiment would require
 polarised antiprotons, but excellent physics, namely
 regarding transversity, can be performed also making
 use of an unpolarised antiproton beam and of a polarised
 protons.

 Also the investigation of the spin dependent cross-sections
 in exclusive ($p\bar{p} \rw \Lambda \bar{\Lambda}$) or
 semi-inclusive ($p\bar{p} \rw \Lambda \bar{\Lambda}X $)
 strange hadrons production (Fig.~\ref{fig:dy-ppss}~right)
 allows for the extraction of the quark distribution and
 fragmentation functions. The correlation between the s
 and $\bar{s}$ quarks can be determined, assuming the
 spin orientation of the $\Lambda$ ($\bar{\Lambda}$)
 given by that of the $s$ ($\bar{s}$) quark, measuring,
 on an event per event basis, the correlation between
 the $\Lambda$ and $\bar{\Lambda}$ polarisations, where
 the formers can be determined studying the angular
 distribution of the decay $p$ ($\bar{p}$) in the self
 analysing weak decay $\Lambda \rw p\pi^-$
 ($\bar{\Lambda} \rw p\pi^-$).

 $\lam$'s and $\alam$'s detection could allow also the
 investigation of open-charm production in antiproton-proton
 scattering 
 $ \bar p p\rw  \Lambda_{c}^+\, X$ .

 Spin dependent measurements will also allow to disentangle
 the electric and magnetic part of the electromagnetic
 form factors in the exclusive dilepton production
 from $p\bar{p}$ annihilation.

 \section{The physics}
 \label{sec:phys}

 \subsection{Parton distribution functions}
 At leading twist, in the case of
 collinear quarks inside the nucleon, or integrating over the
 transverse momentum of the quarks, the quark structure of
 the nucleon is completely described by three distribution
 functions: the unpolarised distribution $f_1(x)$,
 describing the probability of finding a quark with
 a fraction x of the longitudinal momentum of the
 parent hadron, regardless of its spin orientation;
 the longitudinal polarisation distribution $g_1(x)$,
 describing the difference between the number density
 of the quarks with spin parallel and anti-parallel
 to the spin of a parent longitudinally polarised hadron;
 and $h_1(x)$ similar to $g_1(x)$, but  for transverse
 polarisation.


 If we admit a nonzero quark transverse momentum $\kperp$
 and we do not integrate anymore on it, the nucleon
 structure is described, at twist two and three, by eight
 parton distribution functions, among which there are
 some $\kperp$-dependent functions. We'll focus later on
 two $\kperp$-dependent distributions:
 $f^{\perp}_{\rm 1T}(x,\kperp^2)$ and $h^{\perp}_{\rm 1T}(x,\kperp^2)$,
 respectively the distribution functions of an unpolarised
 quark in a transversely polarised hadron, and of a
 transversely polarised quark inside an unpolarised parent hadron.

 For a complete description of hadron production processes,
 the fragmentation functions are needed as well; they
 describe the probability for a quark, in a given
 polarisation state, to fragment into an hadron carrying
 some momentum fraction~$z$.

 A complete review of the theoretical and experimental
 aspects relative to parton distribution and fragmentation
 functions can be found in \cite{bar02}.

 \subsection{Drell--Yan processes}
 Let us focus on the production of muon pairs according
 to the diagram of Fig.~\ref{fig:dy-fey}~left in the
 process $p\bar{p} \rw \mu^+\mu^-X$. Since the virtual
 photon comes from the quark annihilation vertex, any
 asymmetry that can be determined depends on the
 quark distribution functions only functions. 
 
 \begin{figure}[h] 
 \bc
 \begin{minipage}[c]{.48\textwidth}
 {\includegraphics[width=\textwidth]{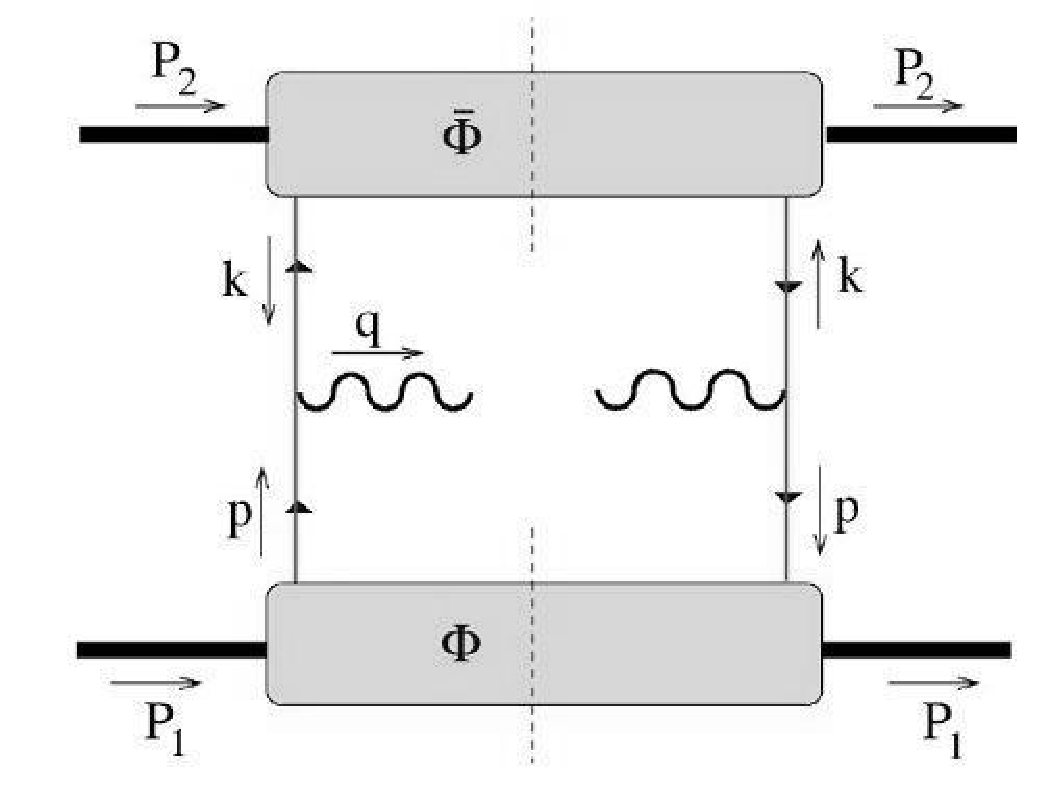}}
 \end{minipage}\hfill
 \begin{minipage}[c]{.48\textwidth}
   {\includegraphics[width=\textwidth]{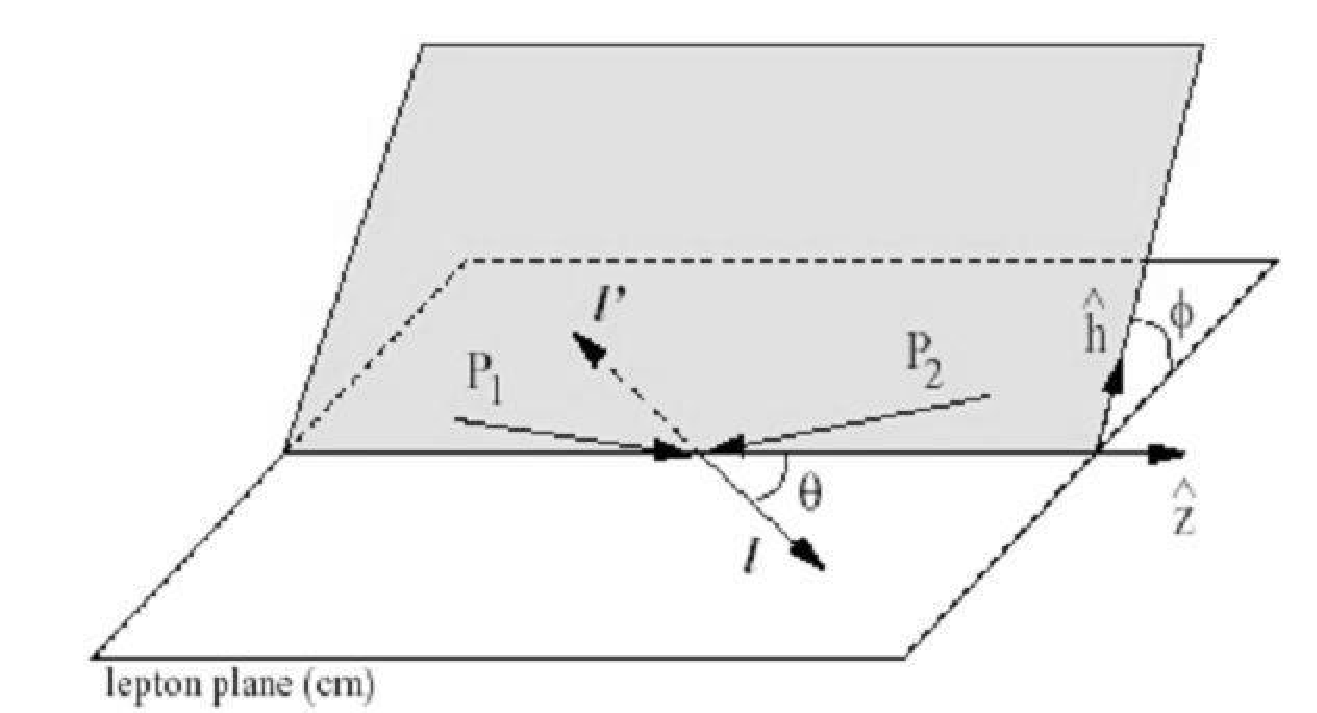}}
 \end{minipage}
 \ec
 \vspace{-6mm}
 \caption{Left: Hand-bag diagram for a Drell--Yan process;
 right: the geometry of the Drell--Yan production in the rest
 frame of the lepton pair  \cite{col77}.}
 \label{fig:dy-hb}
 \label{fig:dy-angles}
 \end{figure}

 The possibility to access chirally odd parton
 distribution functions is probably one of the best
 benefit of the Drell--Yan processes; in such processes
 in fact the quark lines in the diagram (Fig.~\ref{fig:dy-hb}~left)
 are uncorrelated, thanks to the two non-perturbative
 vertexes. Chirally odd amplitudes, and hence
 transversity $h_1(x)$, can be investigated without
 the chiral suppression proper of DIS.



 We will assume herewith the geometry
 (Fig.~\ref{fig:dy-angles}~right) and the kinematic
 variables defined in \cite{col77}. The cross-section
 of the Drell--Yan process $p\bar{p} \rw \mu^+\mu^-X$
 for a given dimuon mass M, is:
 \begin{equation} \label{eq:xs}
 \frac{d^2 \sigma}{d M^2 dx_F} = \frac{4 \pi \alpha^2}{9 M^2 s}
 \frac{1}{(x_1 + x_2)} \sum_{a} e^{2}_a \, \left [f^{a}(x_1)
 f^{\bar{a}} (x_2) \; + \;f^{\bar{a}} (x_1) f^a (x_2)\right ]
 \end{equation}
 being $x_{1,2} = \frac{M^2}{2P_{1,2}q}$ the fractions
 of the longitudinal momenta of the incoming hadrons
 carried by the quark and anti-quarks taking part to
 the annihilation in the virtual photon; the Feynman
 variable $x_f = x_1 - x_2 $, the ratio of the
 longitudinal momentum of the pair to the maximum
 allowable longitudinal momentum in the colliding
 hadrons center of mass frame; the parameter $\tau
 = x_1x_2 = \frac{M^2}{s}$; and the summing is on
 the flavour $a$ of the quark ($a=u,d,{\rm s}$).


 \begin{figure}[h!] 
 \begin{center}
   {\includegraphics[width=.8\textwidth]{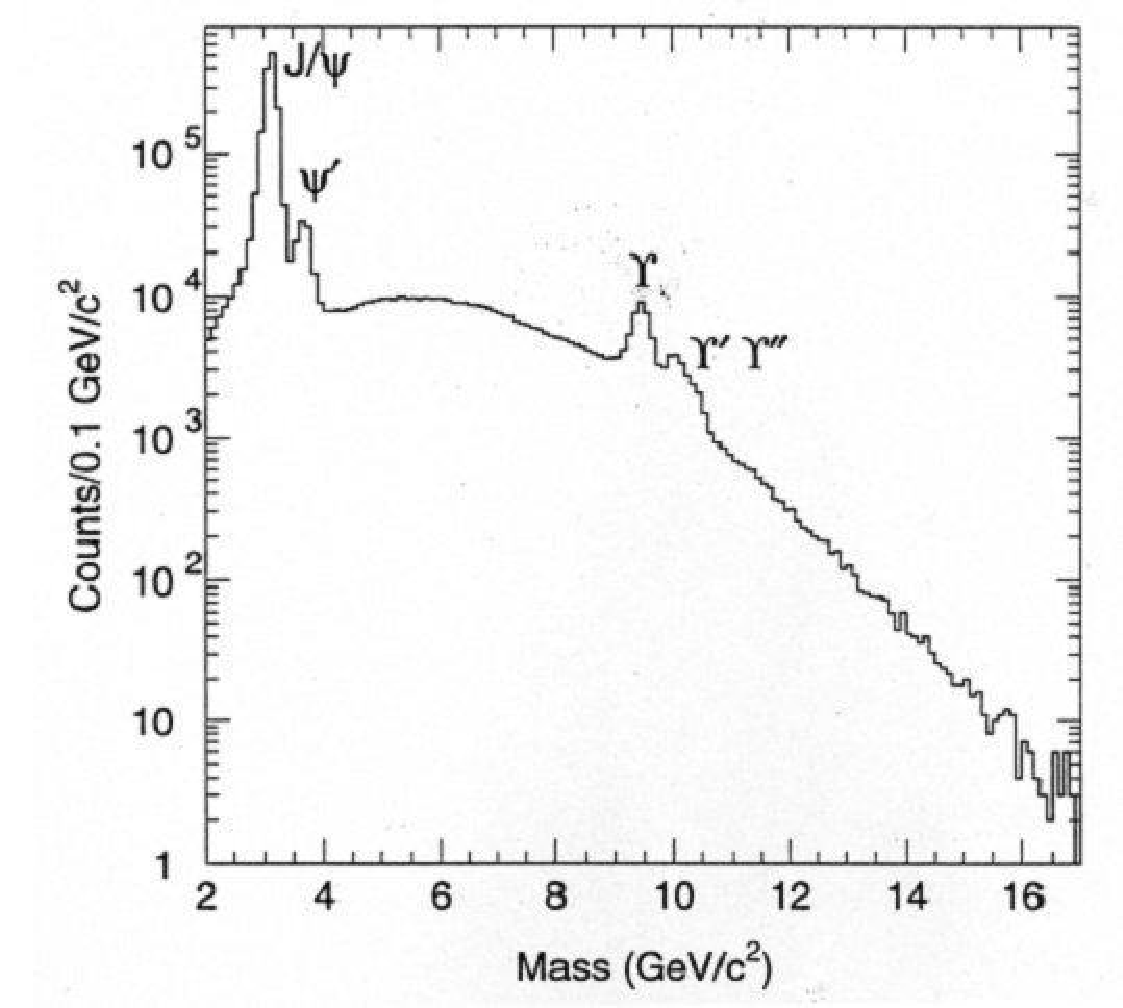}}
 \end{center}
 \vspace{-8mm}
 \caption{Combined dimuon mass spectrum from $pp$ and $pd$
 collisions (from \cite{haw98})}.
 \label{fig:dy-xs}
 \end{figure}

 The scaling properties and the kinematic behaviour
 of $\ppbar \rw \mumu X$ reaction are the same as for
 the $pp \rw \mumu X$; the Drell--Yan cross-section
 \cite{mcgau99,haw98} scales as $d^2 \sigma / d
 \sqrt{\tau} dx_F \varpropto 1 /$s, increasing
 the statistics in the low beam energy region
 consistent with the selection of a di-muon mass in
 the ``safe'' region, i.e. corresponding to values of
 M ranging from 4 to 9\,\gevcc~(Fig. \ref{fig:dy-xs}).
 In the ``safe'' region the dimuon spectrum is
 essentially continuum without resonance effects from
 $ J / \Psi$ and $\Upsilon$ resonance families to
 disentangle in the data analysis. For the data
 arising from the region below the $ J / \Psi$ resonance
 families, perturbative contributions can be important,
 and the formul\ae~that we present later on have to be
 corrected by additional terms; this is the reason why
 in the ASSIA LOI  \cite{assialoi} the safe region only
 is considered to extract the parton distribution
 functions. Nevertheless, since there exists arguments
 in the favour of possibility of studying spin effects
 in the $ J / \Psi$ region  \cite{mauropriv},
 and we intend to investigate also the perturbative
 corrections in the kinematic region below the safe
 region, data in this kinematic region will be collected
 as well. The importance of this perturbative effects
 decreases with increasing $s$ \cite{gam05}.

\begin{figure}[h] 
 \begin{center}
   {\includegraphics[width=.95\textwidth]{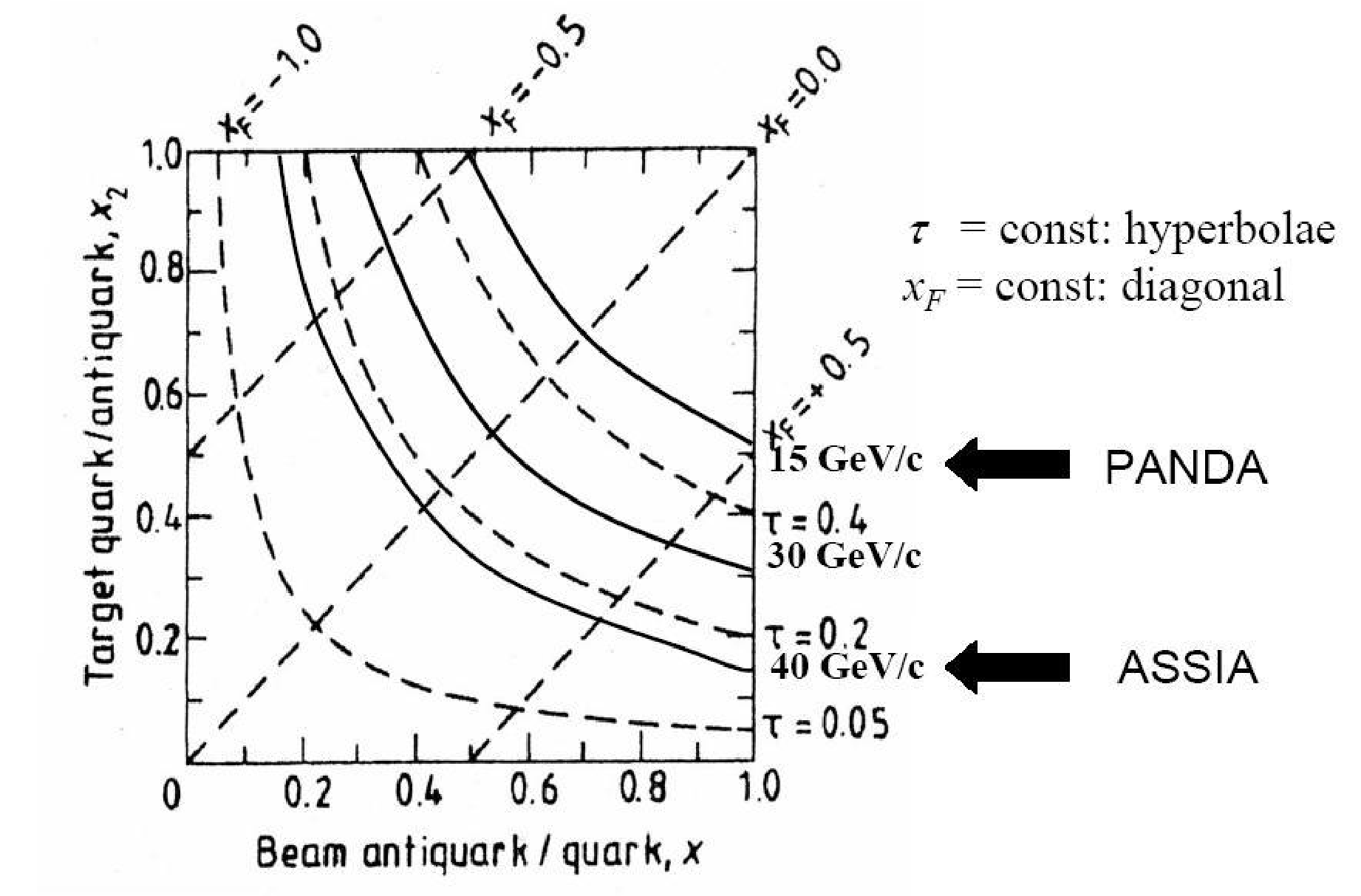}}
 \end{center}
 \vspace{-8mm}
 \caption{Allowed kinematic region in $x_1$ and $x_2$
 for Drell--Yan processes: the regions above the
 hyperbola correspond to the cut on the dilepton mass
 $M>4$\,\gevcc for three different energies of the beam,
 the lowest foreseen for PANDA at HESR, the highest
 proposed by ASSIA for antiprotons extracted from SIS 300.}
 \label{fig:dy-x1x2}
 \end{figure}

 It is important to investigate the parton distribution
 functions in the wide region of $x$;
 a wide $x_1$, $x_2$ region means asking for the $\tau$
 parameter to range from 0~to~1. Moreover, to enlarge the
 statistics, data from the complete safe region should be
 collected. The upper limit of 9\,\gevcc~for M defines
 the highest center of mass energy needed for the complete
 $\tau$ region; the corresponding value for the lowest
 momentum of the $\pbar$ beam is 40\,\gevc. Also rejecting
 the events below the $ J / \Psi$ peak means to cut the
 allowed kinematic region. 
 The allowed region in the scatter plot of the two momenta
 fractions $x_1$ and $x_2$, after asking for a dilepton
 mass above the $ J / \Psi$ peak, i.e. above 4\,\gevcc,
 depends on the $\tau$ value, and thus on center of mass
 energy s. The hyperbola of Fig.~\ref{fig:dy-x1x2} show the
 $\tau$ region selected from the cut on the lowest side of
 the safe region for different values of $s$ and thus of
 the beam momentum. The kinematic region that can be
 accessed making use of antiprotons extracted from SIS 300
 at 40\,\gevc~is wider then the region that could be
 explored by mean of antiprotons colliding on a fixed
 target in the HESR facility, if the beam energy would be
 the one foreseen for PANDA \cite{pandaprop}.

 To ask for the kinetic energy of the $\pbar$ beam the
 value of 40\,\gevc is a reasonable compromise between
 the scaling behaviour of the cross-section and the need
 to cover the wide parton distribution functions range.

 The ideal tool would be a beam and a target both polarised
 either longitudinally or transversely; in such a case the
 following asymmetries could be observed:
 \begin{eqnarray} \label{eq:all}
 A_{\rm LL} &=& \frac{\sum_a e^{2}_{a} g_{1}^a (x_1)
 g_{1}^{\bar{a}}(x_2)}{\sum_a e^{2}_{a} f_{1}^a (x_1)
 f_{1}^{\bar{a}}(x_2)}\\
 \label{eq:att}
 A_{\rm TT} &=& \frac{sin^2 \, \theta \cos \, 2 \phi}{1 \,
 + \cos^2 \theta} \frac{\sum_a e^{2}_{a} h_{1}^a (x_1)
 h_{1}^{\bar{a}}(x_2)}{\sum_a e^{2}_{a} f_{1}^a (x_1)
 f_{1}^{\bar{a}}(x_2)}\\
 \label{eq:alt}
 A_{\rm LT} &=& \frac{2 \sin \,2 \theta \cos\,
 \phi}{1 + \cos^2 \theta} \frac{M}{\sqrt{Q^2}}  
 \frac{\sum_a e^{2}_{a}\left (g_{1}^a (x_1) y g_{\rm T}^{\bar{a}}(_2)
 - x h_{\rm L}^a (x_1) h_{1}^{\bar{a}} (x_2)\right )}
 {\sum_a e^{2}_{a} f_{1}^a (x_1) f_{1}^{\bar{a}}(x_2)}
 \end{eqnarray} 
 where the first asymmetry correspond to both a target and
 a beam longitudinally polarised, the second one to both
 a target and a beam transversely polarised, and the third
 asymmetry to the case of one longitudinally polarised and
 the other transversely polarised. The polar angle $\theta$
 and the azimuthal angle $\phi$ are the ones defined in
 Fig.~\ref{fig:dy-angles}~left and in \cite{col77}.

 It is important to stress that the validity of the formula
 reported above strongly depends on the assumptions that
 the center of mass energy and the Q$^2$ are large enough.

 To extract the parton distribution functions these
 asymmetries have to be compared in a fitting procedure
 to the experimental asymmetries determined at the same
 value of $x_1$ and $x_2$, after the correction by the
 factor $1/(P_{\rm b} \cdot f \cdot P_{\rm T})$
 accounting for the beam polarisation $P_{\rm b}$, the
 dilution factor $f$ and the target polarisation
 $P_{\rm T}$. For an NH$_3$ polarised target, as
 explained in Sec. \ref{sec:det}, the dilution factor,
 that is the number of polarised nucleons over the total
 number of nucleons in the target, would be $f=1/17=0.176$,
 while a target polarisation $P_{\rm T}=0.85$ could be reached.


 These asymmetries could be investigated also at RICH, but
 in the case of a proton-proton scattering, only the sea
 anti-quark would contribute to the Drell--Yan diagram;
 moreover, due to the large center of mass energy, $\sqrt{s}
 \approx 100$\,GeV, the data would be affected by a strongly
 reduced cross section and by quite a small allowed kinematic
 range. The value of the asymmetries itself would be reduced
 as well, due to the much slower evolution of $h_1(x)$ on
 Q$^2$ compared with that of the unpolarised distribution
 functions \cite{bar97}; the numerator of the asymmetry
 ratio would grow slowerly than the denominator, leading thus
 to suppression of the asymmetries for large values of the
 center of mass energy.

 This would not be the case at GSI, where the center of mass
 energy would not be so large, and a wide kinematic region
 could be accessed. If we focus on the double spin asymmetry
 $A_{\rm TT}$, the distribution function of a quark of flavour
 $a$ in the proton can be assumed equal to the distribution
 function of the anti-quark $\bar{a}$ in the antiproton, since
 one can be obtained from the other through a charge conjugation.
 $A_{\rm TT}$ would thus allow a direct access to $h_1(x)$
 squared for the valence quark: $h_{1,q_V}(x_1)h_{1,q_V}(x_2)$.
 It has been shown \cite{ans04} how this asymmetry is expected
 to be huge, $\approx 30\,\%$ for a center of mass energy just
 slightly smaller ($s = 30$--45\,\gevcc).

 Although the availability of both a polarised beam and
 a polarised target is the ideal case, spin effects can also
 be investigated with a polarised target only, or even in
 a completely unpolarised case.

 The angular distribution for dilepton production, for
 unpolarised beam and target is:
 \begin{equation} \label{eq:xs_unp}
 \frac{1}{\sigma} \frac{d \sigma}{d \Omega} =
 \frac{3}{4 \pi} \frac{1}{\lambda + 3} \times
 \left (1 + \lambda \cos^2~\theta + \mu \sin^2 \theta
 \cos \phi + \frac{\nu}{2} \sin^2 \theta \cos 2 \phi\right )
 \end{equation}
 where $\theta$ and $\phi$ are the angular variables above
 defined. Perturbative QCD calculations at next-to leading
 order give $\lambda \approx 1$, $\mu \approx 0$, $\nu \approx 0$,
 confirming the characteristic $\cos^2 \theta$ distribution of
 the decay of a transversely polarised virtual photon, given
 in the parton model; hence, once accounted for the acceptance,
 the experimental cross-section should not depend on the
 azimuthal angle. However fits of experimental data
 \cite{anas88,con89} show remarkably large values 30\,\% of
 $\nu$ at transverse momenta of the lepton pair between 2
 and 3\,GeV. Recently \cite{boer03,col03} it has been pointed
 out that initial state interaction in the unpolarised Drell--Yan
 process could explain the observed asymmetries and be connected
 with the quark (anti-quark) T-odd distributions $h_{1q}^{\perp}$
 and $h_{1\bar{q}}^{\perp}$.

 The measurement, for the $ \bar{p}p \rw \mu^+ \mu^- X$ process,
 of the $\cos 2 \phi$ contribution to the angular distribution
 of the dimuon pair provides the product $h^{\perp}_{1}(x_2,
 \boldsymbol{\kappa}^{2}_{\perp}) \,
 \bar{h}^{\perp}_{1}(x_1,$ $\boldsymbol{\kappa}^{' 2}_{\perp})$.
 This asymmetry can be evaluated also by mean of the PANDA
 detector, where a polarised target cannot be installed because
 of the disturbance of the magnetic field of the solenoid;
 however the maximal antiproton beam energy foreseen for PANDA
 at HESR limits considerably the reachable kinematic domain for
 the Bjorken $x$ variable.

 In the case of a transversely polarised hydrogen target, the
 measured asymmetry for the two target spin states depends on
 the $\sin (\phi \; + \; \phi_{S_1})$ term, where $\phi_{S_1}$
 is the azimuthal angle of the target spin in the frame of
 Fig.~\ref{fig:dy-angles}~right. This term is $\propto h_1(x_2,
 \boldsymbol{\kappa}_{\perp}^2 ) \, \bar{h}_{1}^{\perp} (x_1,
 \boldsymbol{\kappa}_{\perp}^{' 2})$, as shown by \cite{boer99}:

 \begin{eqnarray} \label{eq:at}
 A_{\rm T} &=& \mid \boldsymbol{S_{\perp}} \mid\frac{2 
 \,\sin 2\theta \sin \,( \phi - \phi_{S_1})}{1 + \cos^2 \theta}
 \frac{M}{\sqrt{Q^2}}  \nonumber \\
 &&\frac{\sum_a e^{2}_{a}\left [x \left (f_{1}^{a \perp}(x_1)
 f_{1}^{\bar{a}}(x_2) + y h_{1}^a (x_1) h_{1}^{\bar{a} \perp}
 (x_2)\right )\right ]}
 {\sum_a e^{2}_{a} f_{1}^a (x_1) f_{1}^{\bar{a}}(x_2)}
 \end{eqnarray} 


 The ideal scenario would be to combine double spin measurements
 near the maximum value of the parton distribution functions
 with the investigation of single spin asymmetries as a function
 of the Bjorken $ x$ to evaluate the $x$-dependence of the
 $h_1(X)$ function \cite{bia05}.


 With unpolarised antiprotons and polarised protons, the
 dependence of the quark distribution functions on the quark
 transverse momentum $\kperp$ could be investigated. In
 particular, the measurement of the single spin asymmetry
 (Eq. \ref{eq:at}), in the absence of a polarised beam, is
 a unique tool to probe the $\boldsymbol{\kappa}_{\perp}$ effects.
 Recently, several papers have stressed the importance of
 measuring SSA in Drell--Yan processes [17--21]; 
 these measurements allow the determination of new non
 perturbative spin properties of the proton, like the Sivers
 function, which describes the azimuthal distribution of
 quarks in a transversely polarised proton \cite{ans03}.


 The study of $\bar p^{\uparrow} \, p \to \mu^+ \, \mu^- \, X$
 processes at GSI offers then unique possibilities.

 \subsection{Spin asymmetries in hyperon production}

 It is well known that inclusively produced $\lam'$s, in
 unpolarised $p p$ interactions show a negative transverse
 polarisation, that rises with $x_F$ and $p_{\rm T}$ and
 achieves~40\,\%. Even higher $\lam$'s polarisation
 (60\,\%) was obtained in exclusive reactions like $pp
 \rw p \Lambda K^+$, $pp \rightarrow p \Lambda K^+ \pi^+
 \pi^-$ {\it etc.}~\cite{r608,fel99}. This phenomenon has
 been  confirmed many times in extensive set of experiments;
 yet, its theoretical explanation still remains a persisting
 problem.

 There is a complete correlation between the spin orientations
 of the $\Lambda$ and of the $\bar{\Lambda}$.
 These decays have both a large asymmetry parameter ($\alpha =0.642$)
 and branching ratio (${\rm B.\,R.}=0.640$). 

 Different quark-parton models using static SU(6) wave
 functions were proposed to interpret these polarisation
 effects by introducing a spin dependence into the partonic
 fragmentation and recombination processes~\cite{deg81,and79,dhar90}.
 The $\lam$ polarisation is attributed to some mechanism,
 based on semi-classical arguments~\cite{deg81,and79} or
 inspired by QCD~\cite{dhar90}, by which produced strange
 quarks acquire a large negative polarisation. Recently
 a new approach to this problem based on perturbative QCD
 and its factorisation theorems, and which includes spin
 and transverse momentum of hadrons in the quark fragmentation,
 was proposed in~\cite{ans01}. These models are based on
 different assumptions and are able to explain the main
 features of the $\lam$ polarisation in unpolarised
 $pp$-collisions. To better distinguish between these
 models more complex phenomena have to be considered.

 If the beam or the target is polarised, other observables
 can be accessed, namely the analysing power,
 $A_{\rm N}$\footnote{$A_{\rm N}$ was studied for
 $\pi$-production also, see Sec. \ref{sec:ssa}.}, and the
 depolarisation (sometime referred as spin transfer
 coefficient), $D_{\rm NN}$:

 \begin{eqnarray}
 \label{eq:an}
 A_{\rm N} &=& \frac{1}{P_{\rm B} \cos\phi} \frac{N_{\uparrow}(\phi)
 - N_{\downarrow}(\phi)}{N_{\uparrow}(\phi) +
 N_{\downarrow}(\phi)}, \\
 \label{eq:dnn}
 D_{\rm NN} &=& \frac{1}{2P_B \cos\phi} \left [P_{\Lambda\uparrow}
 \left (1+P_{\rm B} A_{\rm N} \cos(\phi)) - P_{\Lambda\downarrow}
(1-P_{\rm B} A_{\rm N} \cos(\phi)\right )\right ],
 \end{eqnarray}
 where the azimuthal angle $\phi$ is that between the beam
 polarisation direction and the normal to $\Lambda$ production plane.

 It is interesting to note that whereas produced £$\lam$
 polarisation remains large and negative for exclusive and
 inclusive channels the spin transfer coefficient is
 negative in low energy (beam momentum 3.67\,GeV/c) exclusive
 production \cite{disto}, compatible with zero at intermediate
 energies (beam momentum 13.3 and 18.5\,GeV/c) \cite{bon87}
 and positive at high energy (beam momentum 200\,GeV/c)
 inclusive reaction \cite{bra97}. Thus, the measurements
 at 40\,GeV/c can bring an additional information on
 this phenomenon.

 The spin dependence of exclusive annihilation reaction
 $\bar p p \rw \bar \Lambda \Lambda$ has been considered
 as relevant to the problem of the intrinsic strangeness
 component of nucleon \cite{alb95, pak99}. It was
 demonstrated \cite{pas00} that the use of a transversely
 polarised target, in principle, allows the complete
 determination of the spin structure of the reaction.
 Corresponding measurements was performed by PS185
 Collaboration, see \cite{ps185} and references therein.
 Competing models such as $t$-channel meson exchange model
 and $s$-channel constituent quark model reasonably well
 describing the cross-section of this reaction exist. But
 both are unable to describe such spin observables as spin
 transfer from polarised proton to $\lam$ ($D_{\rm NN}$)
 and to $\alam$ ($K_{\rm NN}$). It is evident that new
 data on spin transfer and correlation coefficients at
 higher energies and momentum transfer will be easier
 interpreted in QCD based approaches and can help a better
 understanding the spin dynamic of strong interactions.

 The semi-inclusive $\bar p p \rightarrow \bar
 \Lambda\Lambda X$ production is particularly interesting,
 for which the fundamental Feynman diagram in 
 Fig.~\ref{fig:dy-ppss}~right is relevant; 
 the corresponding diagram for this fundamental process
 would be the one of Fig.~\ref{fig:dy-hb}~left with the
 virtual photon replaced by a gluon. Therefore the two
 quarks chiralities are unrelated and there is not a chiral
 suppression of $h_1(x)$, like in DIS.

 Therefore, even with an unpolarised antiproton beam but
 with a polarised target one can get the spin correlation
 parameters related both to the parton distributions and
 to the quark fragmentation functions.

 \subsection{Single spin asymmetries}
 \label{sec:ssa}

 Besides the single spin asymmetry defined in Eq. \ref{eq:at}
 for the Drell--Yan processes, also the investigation of
 spin effects in hadron production by the mean of the single
 spin asymmetry:
 \begin{equation}
 \label{eq:an_a}
 A_{\rm N} = \frac{d\sigma^\uparrow - d\sigma^\downarrow}
 {d\sigma^\uparrow + d\sigma^\downarrow} \>,
 \end{equation}
 would be of relevant importance in the framework of a possible
 QCD phenomenology. This kind of single spin asymmetry has been
 measured in $p^{\uparrow} \, p \to \pi \, X$ and $\bar p^{\uparrow}
 \, p \to \pi \, X$ processes, and at large values of $x_F$ ($x_F
 \gtrsim 0.4$) and moderate values of $p_T$ ($0.7 < p_T < 2.0$\,GeV/c)
 have been found by several experiments \cite{bon87,ada91,ada04}
 to be unexpectedly large. The pion production at large $x_F$
 values originates from valence quarks, and according to \cite{ans05}
 the $A_{\rm N}$ behaviour (positive for $\pi^+$ and negative
 for $\pi^-$) can be explained by Sivers effect, while Collins
 effect gives negligible contribution; 
 similar values and trends of $A_{\rm N}$ have been found in
 experiments with center of mass energies ranging from 6.6 up
 to 200\,GeV: this seems to hint at an origin of $A_{\rm N}$
 related to fundamental properties of quark distribution
 and/or fragmentation.

 A new experiment with anti-protons scattered off polarised
 protons, in a new kinematic region, could certainly add
 information on such spin properties of QCD. Also, $A_{\rm N}$
 observed in $\bar p \, p^{\uparrow} \to \pi \, X$ processes
 should be related to $A_{\rm N}$ observed in $\bar p^{\uparrow}
 \, p \to \pi \, X$ reactions, which should be checked.

 \subsection{Electromagnetic form factors}

 The study of nucleon electromagnetic form factors is
 a powerful tool to investigate the nucleon structure;
 in particular the form factors in the time-like region,
 which can be measured through the reactions $\overline{p}
 +p\leftrightarrow e^+ +e^-$, provide additional
 informations on the nucleon structure with respect to
 the ones that can be obtained by the mean of eN-scattering
 in the space like region.
 The two regions, space-like and time-like, can be
 connected analytically through dispersion relations.
 The form factors data available extend mostly in the
 space-like region; the data in the time-like region
 reach higher Q$^2$, but are less precise; low statistics,
 imprecise measurements for cross-section (and only for
 protons and not for neutrons), and no spin effects in the
 time-like region data are investigated.

 The reaction $\bar{p} p \rw \mu^+ \mu^-$ with polarised
 $p$ ($\bar{p}$) can be an alternative way to study form
 factors in the time-like region measuring both the
 angular distributions of the differential cross-sections
 and of the analysing power. The angular dependence of
 the differential cross section for $\overline{p}+p\to
 \mu^+ +\mu^-$ can be expressed as a function of the
 angular asymmetry ${\cal R}$:
 \begin{equation}
 \displaystyle\frac{d\sigma}{d(\cos\theta)}=
 \sigma_0\left [ 1+{\cal R} \cos^2\theta \right ],~
 {\cal R}=\displaystyle\frac{\tau|G_M|^2-|G_E|^2}{\tau|G_M|^2+|G_E|^2}
 \label{eq:eq3}
 \end{equation}
 being $\sigma_0 = \sigma_{\theta=\pi/2}$.

 Theoretical models provide for these quantities predictions
 very sensitive on the different underlying assumptions on
 the $s$-dependence of the form factors. Moreover
 polarisation effects are particularly interesting, since
 a transverse polarisation $P_{\rm T}$ (of the proton or of
 the antiproton) results in a nonzero analysing power:
 \begin{eqnarray}
 {\cal A}&=&\displaystyle\frac{\sin 2\theta Im G_E^*G_M}{D\sqrt{\tau}},
 ~D=|G_M|^2(1+\cos^2\theta)+\displaystyle\frac{1}{\tau}|G_E|^2\sin^2\theta\\
 \displaystyle\frac{d\sigma}{d\Omega}(P_{\rm T})&=&
 \left ( \displaystyle\frac{d\sigma}{d\Omega} \right )_0 \left [1+{\cal A}P_{\rm T}
 \right ]
 \end{eqnarray}
 The $\tau$-dependence of ${\cal A}$ is very sensitive to
 existing models of the nucleon form factors, which
 reproduce equally well the data in the space-like
 region.

 Therefore, a precise measurement, either with a polarised
 antiproton beam or with an unpolarised antiproton beam on
 polarised protons, would be very interesting in order to
 achieve a global interpretation of the four nucleon form
 factors, electric and magnetic, for proton and neutron,
 both in the space-like and in the time-like region. The
 angular distribution of the produced leptons for the
 channels $\bar{p} p \rw l^+ l^-$ allows for the separation
 of the electric and magnetic form factors, since the
 asymmetry ${\cal R}$ is sensitive to the relative value
 of $G_M$ and $G_E$.  Moreover, an experimental proof of
 a large relative phase of proton  magnetic and electric
 form factors at relatively large momentum transfers in
 the time-like region would be a strong indication of
 a different behaviour of these form factors.

 \subsection{Open charm}
\begin{figure}[h!] 
 \begin{center}
 {\includegraphics[width=.8\textwidth]{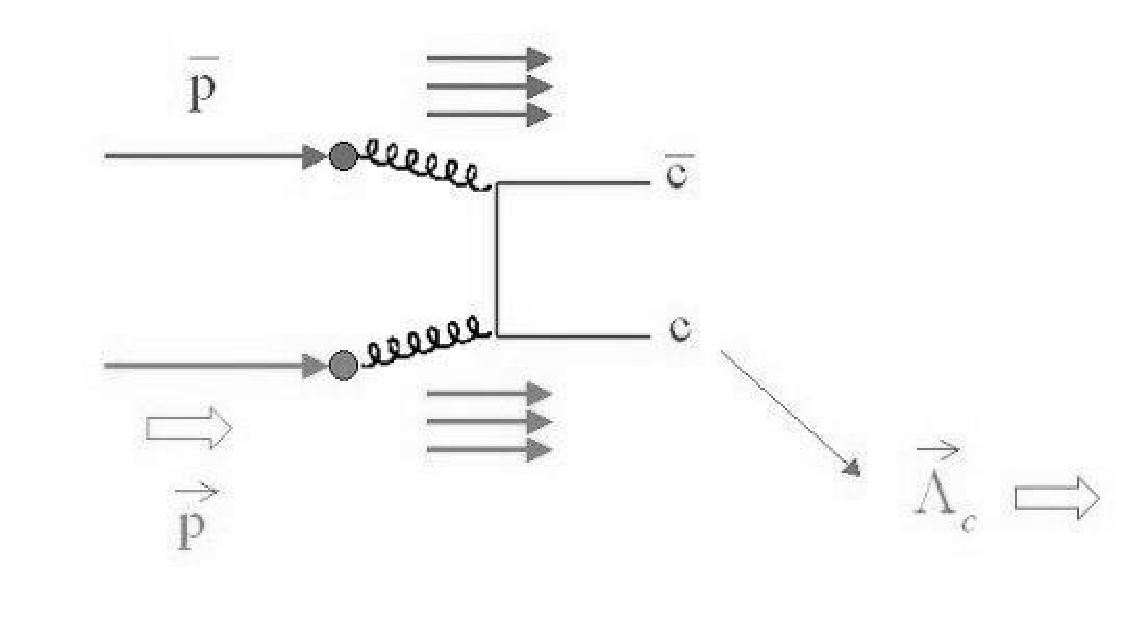}}
 \end{center}
 \vspace{-10mm}
 \caption{Open charm production with antiproton beams.}
 \label{fig:openc}
 \end{figure}

 Very recently the possibility to use the antiproton
 beam for open charm production (Fig. \ref{fig:openc})
 has been suggested \cite{bropriv}. In this process
 both $\it{g g}$ and $q \bar q$ fusion play an
 essential role in different kinematic regions and
 could be used to probe the internal structure of
 the nucleon. For both, the antiproton beam is an
 ideal probe. This study can be pursued detecting, for
 example, the $ \bar p p \rw \Lambda_{c}^+\, X$
 production, where the $\Lambda_{c}^+ \rw \Lambda \,
 \pi^+$ weak decay and the $\Lambda_{c}^+ \rw \Lambda
 \, e^+ \, \nu_c $ semileptonic decay can be used
 to infer the c polarisation. The asymmetry parameters
 for these decays are huge but the branching ratios
 small so that the feasibility of this option has to
 be worked out with simulations. It looks very
 attractive as no data exist in the $s \approx
 80$--90\,GeV$^2$ region.

 \section{Experimental set-up}

 \subsection{Beam and target}
 \label{sec:beam}
 The antiproton beam energy foreseen at HESR is 15\,GeV/c,
 with a luminosity, in the case of the PANDA pellet target,
 of $\le 2 \times 10^{32}$\,cm$^{-2}$\,s$^{-1}$ and a momentum
 spread lower than $\pm 1 \times 10^{-4}$; these excellent
 performances do not fit with the experimental program we
 propose, that requires a minimal energy of 40\,GeV and
 a limited momentum resolution. Moreover the present design
 of the PANDA detector \cite{pandaprop} is not compatible
 with a polarised target.

 A different solution could then be foreseen, among the
 two described in details herewith.  

 A first possibility would be an antiproton beam of energy
 $\geq40$\,GeV/c extracted from SIS 300 scattering on
 a polarised target. The expected momentum spread of such
 a beam should be about $\pm 2 \times 10^{-4}$, that is
 largely enough for the proposed measurements. 

 Assuming for the extraction the whole foreseen antiproton
 production accumulation rate of $7 \times 10^{10}\,\bar{p}/
 {\rm h}$, and injection and extraction efficiencies
 always larger than 0.90, the expected beam intensity on
 the target is of about $1.5 \times 10^{7}\,\bar{p}~\cdot$\,s$^{-1}$.
 The target could be similar to the one of
 the COMPASS apparatus at CERN \cite{compprop}, where two
 cells with opposite polarisation are put one downstream
 of the other, a solution that allows for the minimisation
 of the systematic errors in asymmetry measurements. In the
 case of a NH$_3$ target 15\,g/cm$^2$ thick, with a dilution
 factor $f=1/17=0.176$ and a polarisation $P_{\rm T}=0.85$,
 the expected luminosity is:
 \begin{equation}
 \label{eq:lum}
 \boldsymbol{ \mathcal{L}}=\frac{3}{17} \times 15 \times 6
 \cdot 10^{23} \times 1.5 \cdot 10^7 = 2.25 \cdot
 10^{31}\,{\rm cm}^{-2}\,{\rm s}^{-1}
 \end{equation}

 \begin{figure}[h!]
 \begin{center}
   {\includegraphics[width=.935\textwidth]{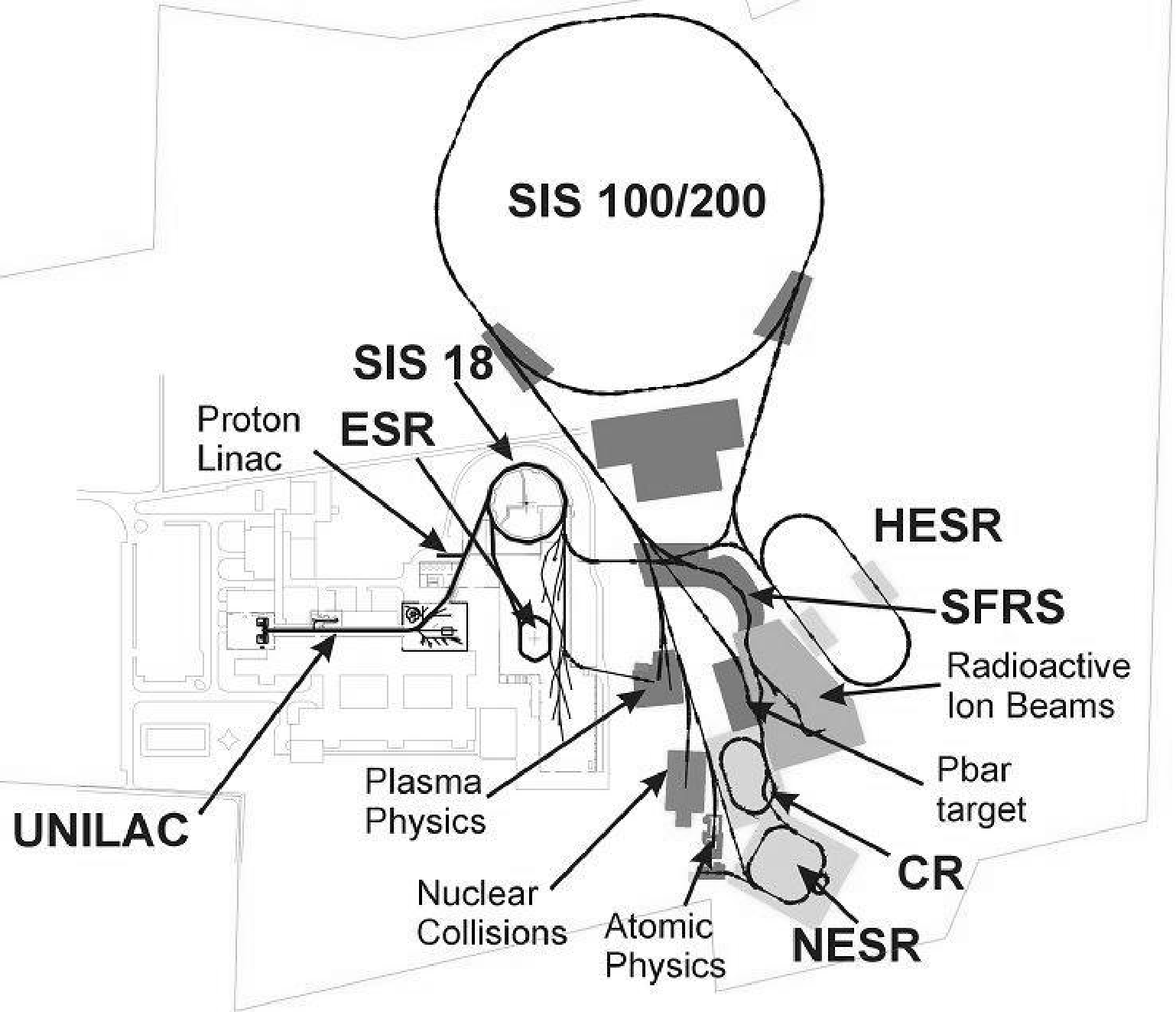}}
 \end{center}
 \vspace{-3mm}
 \caption{ The new facility layout at GSI; SIS 300 would
 be an upgrade of the already foreseen SIS 100/200 synchrotron.}
 \label{fig:gsi}
\end{figure}

 The generation of a 40\,GeV/c antiproton beam would
 require the following additional construction or
 modification items to the presently proposed configuration
 scheme (Fig. \ref{fig:gsi}) of the new International
 Accelerator Facility in GSI:
 \begin{enumerate}
 \item extraction of the accelerated antiproton beam from
 SIS 100 into SIS 300, requiring a transition system to
 be designed and built; or alternatively an injection
 scheme from the CR into the SIS 300;
 \item a slow extraction system from SIS 300 into a more
 powerful extraction beam-line able to handle momenta
 larger than 40\,GeV/c;
 \item a cave housing the experimental setup as proposed
 which can handle the expected radiation doses (with
 $\approx 2 \times 10 ^7 \; \bar{p} \cdot {\rm s}^{-1}$).
 \end{enumerate}
 With this scheme, and provided that different spin
 orientations were available for the polarised target
 like in the COMPASS set-up \cite{compprop}, both
 longitudinal and transverse asymmetries could be
 measured. In addition, if a transversely polarised
 antiproton beam could be produced and extracted from
 SIS 300, a unique tool for the study of the nucleon
 structure would be available.

 An alternative solution, proposed by Hans Gutbrod
 \cite{gut05}, could be to imagine the HESR as
 a collider with both a polarised proton and antiproton
 beams interacting with a luminosity comparable to that
 reachable with an external target; the longitudinal
 size of the antiproton-proton colliding region should
 not be larger than few millimetres. If such a
 luminosity could be reached, the advantage of such
 a solution would be that thre will be no loss of
 accuracy due to diluition factor,
 making the asymmetries measurement more precise for
 the same number of events collected. 
 The required CM energy $\sqrt{s} \approx 30$\,GeV for
 the proposed program could be easily reached with the
 present foreseen performances of HESR (15\,GeV/c). In
 addition the higher CM energies available would allow
 new physics opportunities.

 
 A polarised proton beam of up to 15\,GeV/c would
 require a polarised proton source and an acceleration
 scheme preserving the polarisation. No new beam line
 needs to be built and no additional extraction needs
 to be included into the acceleration system. The
 lattice of the HESR would have to allow an
 interaction region of both beams.

 The key issues of these two proposals is the luminosity. 


 \vspace{1mm}
 \subsection{Detector concept}
 \label{sec:det}
 \vspace{1mm}

\begin{figure}[h!] \vspace{2mm}
 \begin{center}
   {\includegraphics[width=.75\textwidth]{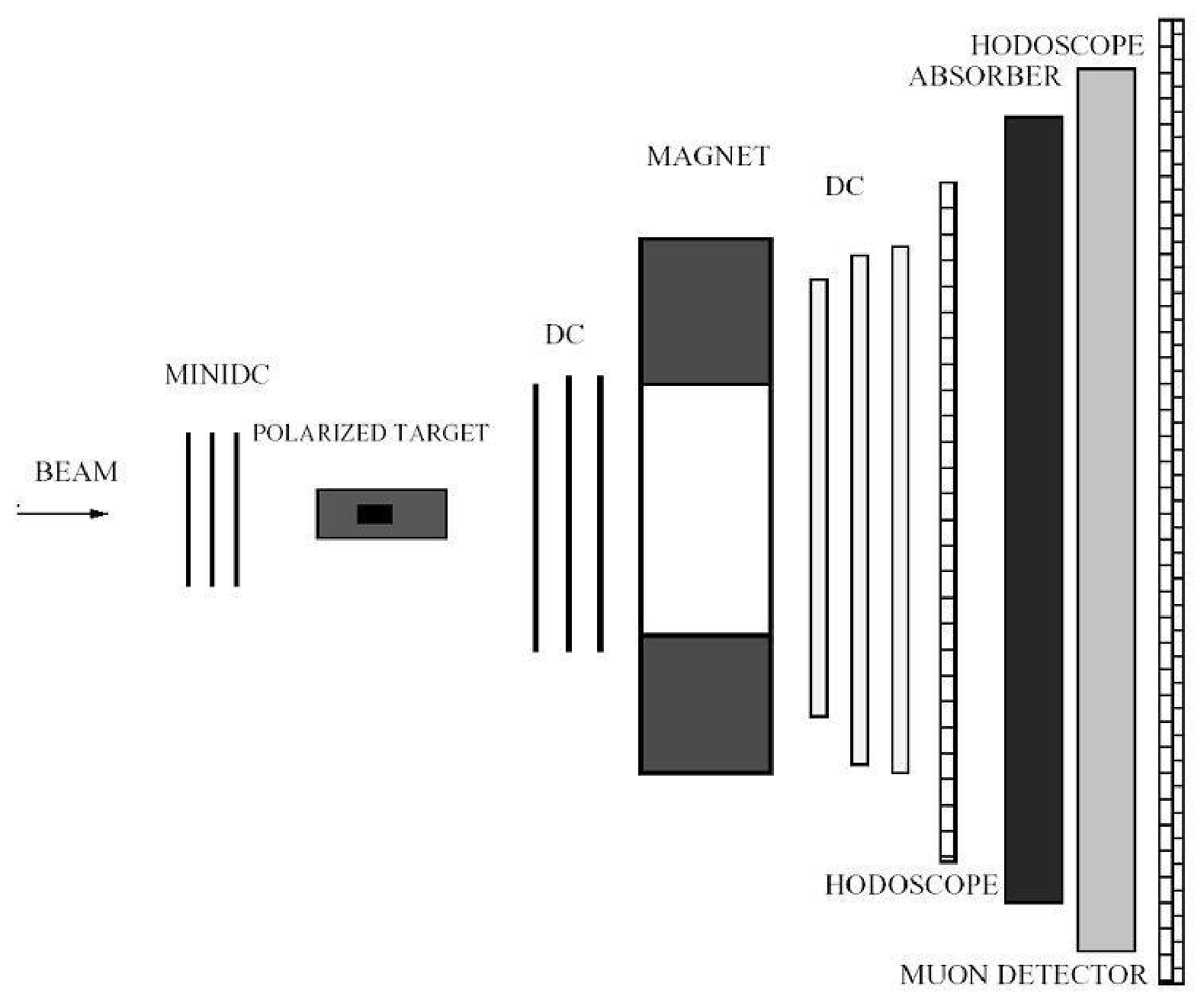}}
 \end{center}
 \vspace{-3mm}
 \caption{ Sketch of the apparatus. MiniDC stay for such
 detectors of drift type like GEM's and $\mu$MEGAs
 in COMPASS. DC Stay for a combination of small drift type
 detectors with high spatial resolution with 
 larger detector with a dead central area. }
 \label{fig:setup}
 \end{figure}

 The proposed detector concept (Fig.~\ref{fig:setup}) is
 inspired from the Large Angle Spectrometer, that is the
 first part of the COMPASS spectrometer \cite{compprop}.
 Such a detector concept would be compatible with an
 extraction scheme of the antiprotons from SIS 300;
 if the HESR collider mode should become available,
 a different set-up, not discussed herewith, should be
 foreseen.

 The Large Angle Spectrometer consists of a large dipole
 magnet SM1 (a window frame magnet with an aperture of
 $2.0 \times 1.6$, a depth of about 1\,m, providing
 a field integral of about 1\,Tm) and tracking detectors
 of different types, chosen in such a way that they can
 sustain the beam rate ($1.5 \times 10^7 \; \bar{p}$/s)
 and provide the hits position with such a precision to
 guarantee the needed resolution for the position of the
 vertexes of the decaying particles ($\Lambda$ and
 $\bar{\Lambda}$) and for the widths of the corresponding
 peaks in the invariant mass spectra. To reach these
 goals and also to minimise the overall cost of the
 apparatus, detectors of smaller size but with thinner
 resolution and accepting higher rates have been chosen
 to detect the hits nearer to the beam trajectory. These
 detectors are GEM and MICROMEGAS, that provide spatial
 resolutions with $\sigma \leq 70\,\mu$. To detect hits
 at larger distances from beam trajectories MWPC and
 STRAW tubes are used that provide spatial resolutions of
 the order of the millimetre. These last detectors have
 a dead zone in their central part, that is nearer to the
 beam trajectories and covered by the GEM and MICROMEGAS
 detectors.

 With this setup a mass resolution ($\sigma \approx
 2.5$\,MeV/c$^2$) can be obtained for the $\Lambda$
 ($\bar{\Lambda}$). The expected spatial resolution on
 the position of the decay vertexes of the $\Lambda$
 ($\bar{\Lambda}$) goes from $\approx 1$\,cm, for very
 small angles with the beam trajectories, to a couple
 of mm for larger angles. This spatial resolution is
 large enough to base the $\Lambda$ ($\bar{\Lambda}$)
 identification on the requirement that these
 vertexes are outside the target.

Trigger is provided by scintillating hodoscopes,
 asking for a multiplicity greater then 2; muon
 detection is performed by the mean of sandwiches
 of iron plates, IAROCCI tubes and scintillator slabs,
 already present in the COMPASS apparatus.

 A vacuum pipe of growing cross-section would catch
 the beam up to the beam dump to minimise the
 background related to the interactions of the beam
 after the target.

 \subsection{Counting rates}
 Since the Drell--Yan channel $\bar{p} p \rw \mu^+
 \mu^- X$ is affected by a low cross section, and the
 hyperon production channel $\bar p +p \rw \bar
 \Lambda+\Lambda + X$ is affected by strong limits in
 acceptance, the expected counting rates for these two
 reactions will be considered as the limit cases; the
 case of the proton time like form factors will be
 discussed later on. An intensity of the antiproton
 extracted beam at the polarised target of $1.5 \times
 10^7\,\bar{p} \cdot$ s$^{-1}$ is assumed, while the
 expected luminosity is the one of Eq. \ref{eq:lum}.

 The expected cross section for the $\bar{p} p
 \rw \mu^+ \mu^- X$ reaction, integrated over positive
 $x_F$ and all transverse momenta, is about 0.3\,nb at
 40\,\gevc; it can be extracted scaling as 1/s the data
 acquired \cite{anas88} for $\ppbar \rw \mumu X$ at
 125\,\gevc for dimuon masses ranging between 4 and
 9\,\gevcc. The expected counting rate is then:
 \begin{equation}
 \boldsymbol{ \mathcal{R}}=2.25 \cdot 10^{31} \times 3
 \cdot 10^{-34} \times A=6.75 \cdot 10^{-3} \times A
 \cong 3 \cdot 10^{-3}\,{\rm ev. \, s}^{-1}
 \end{equation}
 for the acceptance $A=0.44$, consistent with the
 horizontal ($\Delta \theta=\pm\,500$\,mrad) and vertical
 ($\Delta\phi=\pm\,300$\,mrad) acceptances of the
 spectrometer scheme describe in Sec. \ref{sec:det}.

 Assuming for 20 useful hours per day of data taking,
 one gets about 200\,events/ day, that is in two periods
 of 100\,days, 40000 dimuons events in the continuum
 with masses larger than 4\,GeV. Such a statistics,
 remembering the $M$ dependence of the cross-section
 that decreases by about two order of magnitude from
 $M=4$\,GeV to $M=9$\,GeV, will allow to inspect the
 $x \to 1$ region where higher twist contributions are
 expected \cite{brod82}.

 For the reaction $\bar p +p \rw \bar \Lambda+\Lambda + X$
 a cross-section $\sigma=400\,\mu$b is given \cite{bal}.
 Assuming an acceptance $A=0.02$, that accounts for the
 detection of the $\Lambda$ ($\bar \Lambda$) through
 their weak decay $\Lambda \rw p \pi^-$ ($\bar \Lambda
 \rw \bar p \pi^+$), the expected counting rate is:
 \begin{equation}
 \boldsymbol{ \mathcal{R}}=2.25 \cdot 10^{31} \times 4 \cdot
 10^{-28} \times A = 9.0 \cdot 10^{3} \times A \approx
 2 \cdot 10^{2}\,{\rm ev.\,s}^{-1}
 \end{equation}
 that means, within the same assumptions as before,
 $\approx \,1.4 \cdot 10^7$ ev/day.

 In the case of the open charm production, the cross-section for the 
 $\bar p p \rw \Lambda_{c}^+\, X$ at 40\,GeV is
 expected to be of the order of the $\mu$barn
 \cite{vog97}; the detection of the $\lam$'s coming
 from the $\Lambda_{c}^+ \rw \Lambda \,\pi^+$ weak
 decay is affected by the same acceptance as quoted
 above ($A=0.02$) and by the branching ratio of the
 weak decay ($BR_{\Lambda_{c}^+ \rw \Lambda \,\pi^+}=
 0.9$\,\%). The expected counting rate is then:
 \begin{equation}
 \boldsymbol{\mathcal{R}} = 2.25 \cdot 10^{31} \times 1 \cdot
 10^{-30} \times 9 \cdot 10^{-3} \times A = 2.02 \cdot
 10^{-1} \times A \approx 4 \cdot 10^{3}\,{\rm ev.\,s}^{-1}
 \end{equation}
 that means $\approx \,30000$\,ev /100\,days.

 For the proton time like form factors only antiproton
 beams of energyup to 10\,GeV would  be selected,
 because with a luminosity $\boldsymbol {\mathcal{L}}
 = 2 \cdot 10^{32}$ only 2.4 events/day are expected
 at $s = 20$\,GeV$^2$.

 \section{Conclusion}

 The different physics items discussed in Sec.
 \ref{sec:phys} provide excellent tools to deepen the
 investigation of the nucleonic structure; the ideal
 tools would be both polarised antiprotons and protons,
 but even with polarised protons only, plenty of
 spin effects can be investigated as well.

 The key issue is the total energy in the center of
 mass frame $\sqrt{s}$, in order to allow the
 investigation of the parton distribution functions in
 a $x$ Bjorken domain large enough. This goal can be
 achieved in two different ways: either with a slow
 extraction from SIS 300 of an antiproton beam,
 eventually polarised, of momentum greater than 40\,\gevc,
 scattering on a fixed polarised target, that can be
 polarised both transversely and longitudinally;
 or in an HESR modified to a collider of an antiproton
 beam, eventually transversely polarised, and of
 a transversely polarised proton beam.

 In the former scenario more information on spin effects
 could be accessed, since the asymmetries related to
 both the transverse and the longitudinal polarisation
 of the nucleon would be investigated. 

 The latter scenario would have the advantage, for an
 equal luminosity, of a better factor of merit, for
 a proton polarisation equal to that of the target, as
 no dilution factor has to be taken into account in
 that case. A larger acceptance from the detector point
 of view could be obtained, being the cut in acceptance
 for very forward or backward emitted pairs of muons or
 $\lam$'s smaller. Also the modification to the new GSI
 facility layout would be smaller (Sec. \ref{sec:beam}).

 The collider mode, that is under investigation presently,
 has evolved in different scenarios, studied to get the
 highest possible luminosity and including the possibility
 to use the PANDA detector also for the study of spin
 physics. In the case of an asymmetric collider, this
 detector could be used not only for preliminary studies
 of azimuthal asymmetries for unpolarised protons versus
 unpolarised antiprotons at the maximum momentum 15\,GeV/c,
 but also to detect single and double spin asymmetries
 with transversely polarised protons (antiprotons) at
 $\sqrt{s} \ge 10$\,GeV and therefore in a kinematic
 region where perturbative corrections are expected to
 be smaller.

 \bigskip



 \end{document}